\documentstyle[11pt]{article}

\def\fnote#1#2{\begingroup\def\thefootnote{#1}\footnote{#2}\addtocounter
{footnote}{-1}\endgroup}

\begin{document}

\hfill{UTTG-03-06}

\vspace{36pt}

\begin{center}
{\large {\bf {Quantum Contributions to Cosmological Correlations II: Can These Corrections Become Large?}}}

\vspace{36pt}
Steven Weinberg\fnote{*}{Electronic address:
weinberg@physics.utexas.edu}\\
{\em Theory Group, Department of Physics, University of
Texas\\
Austin, TX, 78712}

\vspace{30pt}

\noindent
{\bf Abstract}
\end{center}
\noindent
This is a sequel to a previous detailed study of quantum corrections to cosmological correlations.  It was found there that except in special cases these corrections depend on the whole history of inflation, not just on the behavior of fields at horizon exit.  It is shown here that at least in perturbation theory these corrections can nevertheless not be proportional to positive powers of the Robertson--Walker scale factor, but only at most to powers of its logarithm, and are therefore never large.

 \vfill

\pagebreak
\setcounter{page}{1}

\begin{center}
{\bf I. INTRODUCTION}
\end{center}

Calculations of non-Gaussian corrections to cosmological correlations in the classical approximation have shown that these higher-order corrections are generally suppressed by powers of $GH^2$, where $G$ is Newton's constant and $H$ is the cosmological expansion rate at the time of horizon exit[1].  From the magnitude of the observed Gaussian correlations of fluctuations in the cosmic microwave background, it is known that $GH^2\approx 10^{-12}$ for the fluctuation wavelengths studied in the microwave background, which makes the expected non-Gaussian corrections quite small.  {\em  Quantum} effects involve additional powers of $G$, and are therefore usually supposed to be too small to be detected.  But a recent paper[2] has shown that it in many theories these quantum corrections depend on the whole history of inflation, not just on the behavior of fields at horizon exit, raising the possibility that they may be much larger than usually thought.  In particular, if quantum corrections were to involve positive powers of the Robertson--Walker scale factor $a(t)$ at the end of inflation, then they might be large enough to be detected.  In this paper we will extend the results of reference [2] to a very large class of theories, and show that (at least in perturbation theory) this never happens; quantum corrections depend at most on powers of $\ln a(t)$, and  therefore (without $\approx 10^{12}$ $e$-foldings after horizon exit) never become large.

\begin{center}
{\bf II. SCALARS AND GRAVITATION}
\end{center}

We will first consider a theory of multiple scalar fields $\varphi_n(x)$ and gravitation, slightly more general than that considered in reference [2].  The scalar Lagrangian is assumed to consist of a conventional minimal kinematic term, plus a term with arbitrary potential $V(\varphi)$.  In the Arnowitt-Deser-Misner formalism[3], the components of the metric are
\begin{equation}
g_{ij}=a^2e^{2\zeta}[\exp{\gamma}]_{ij} ,~~~~~\gamma_{ii}=0\;,
\end{equation}
\begin{equation}
g_{00}=-N^2+g_{ij}N^iN^j\;,~~~~~~~g_{i0}=g_{ij}N^j\;,
\end{equation}
where $a(t)$ is the Robertson--Walker scale factor, $\gamma_{ij}({\bf x},t)$ is a gravitational wave amplitude,  $\zeta({\bf x},t)$ is a scalar, and $N$ and $N^i$ are auxiliary fields, whose time-derivatives do not appear in the action.
The Lagrangian density  (with $8\pi G\equiv 1$) is
\begin{eqnarray}
&&{\cal L}=\frac{a^3}{2}e^{3\zeta}\Bigg[-2NV(\varphi)+N^{-1}\Big(E^j{}_iE^i{}_j-(E^i{}_i)^2\Big) \nonumber\\&&~~~~~~+N^{-1}\sum_n\Big(\dot{\varphi}_n-N^i\partial_i\varphi_n\Big)^2\Bigg]-\frac{Na}{2}e^{\zeta}[\exp{(-\gamma)}]^{ij}\sum_n\partial_i\varphi_n\partial_j\varphi_n\nonumber\\&&
~~~~~~+\frac{a}{2}Ne^\zeta \left[\exp(-\gamma)\right]^{ij}\,R^{(3)}_{ij}\;,
\end{eqnarray}
where 
\begin{eqnarray}
&&~~~~~~~E_{ij}\equiv \frac{1}{2}\Big(\dot{g}_{ij}-\nabla_iN_j-\nabla_jN_i\Big)\;,\nonumber\\&&E^i{}_j=a^{-2} e^{-2\zeta}\left[\exp(-\gamma)\right]^{ik}E_{kj}\;,~~N^i=a^{-2} e^{-2\zeta}\left[\exp(-\gamma)\right]^{ik}N_k\;,
\end{eqnarray}
where $\nabla_i$ is the three-dimensional covariant derivative calculated with   the three-metric $e^{2\zeta}\gamma_{ij}$; and $R_{ij}^{(3)}$ is the curvature tensor calculated with this three-metric.
The auxiliary fields $N$ and $N^i$ are to be found by requiring that the action is stationary in these variables.  This gives the constraint equations:
\begin{equation}
\nabla_i\left[N^{-1}\Big(E^i{}_j-\delta^i{}_jE^k{}_k\Big)\right]=N^{-1}\sum_n\partial_j\varphi_n\Big(\dot{\varphi}_n-N^i\partial_i\varphi_n\Big)\;,
\end{equation}
\begin{eqnarray}
&& N^2\left[R^{(3)}-2V-a^{-2}e^{-2\zeta}[\exp(-\gamma)]^{ij}\sum_n\partial_i\varphi_n
\partial_j\varphi_n\right]=E^i{}_jE^j{}_i-\left(E^i{}_i\right)^2\nonumber\\&&~~~~~~~~~~~~
+\sum_n\Big(\dot{\varphi}_n-N^i\partial_i\varphi_n\Big)^2\;.
\end{eqnarray}
For our present purposes, all we need to know about $R^{(3)}_{ij}$, $E^i{}_j$, and the solutions for $N$ and $N^i$ is that none of them contain terms with positive powers of $a$.  
We can impose gauge conditions, for instance by setting any one of the scalar fields equal to its unperturbed value and requiring that $\partial_i\gamma_{ij}=0$.

The possibility of positive powers of $a(t)$ in correlation functions of the $\varphi_n$ and/or $\zeta$ arises from the explicit factors $a^3$ and $a$ in the terms in Eq.~(3).  But these factors can be compensated by negative powers of $a(t)$ in various field time derivatives and in various commutators, which arise from the structure of the perturbative expansion for correlation functions.  The expectation value of any product $Q(t)$ of field operators at various space points (but all at the same time $t$)  is [2]
\begin{eqnarray}
\langle Q(t)\rangle &=&\sum_{N=0}^\infty i^N\, \int_{-\infty}^t dt_N \int_{-\infty}^{t_N} dt_{N-1} \cdots \int_{-\infty}^{t_2} dt_1 \nonumber\\&&\times\left\langle \Big[H_I(t_1),\Big[H_I(t_2),\cdots \Big[H_I(t_N),Q^I(t)\Big]\cdots\Big]\Big]\right\rangle\;,
\end{eqnarray}
(with the $N=0$ term   understood to be just $\langle Q^I(t)\rangle $).  Here $Q^I$ is the product $Q$ in the interaction picture (with time-dependence generated by the part of the Hamiltonian  that is quadratic in fluctuations); and $H_I$ is the interaction part of the Hamiltonian (the part that is of third or higher order in fluctuations) in the interaction picture.  The fields in the interaction picture are 
\begin{equation} 
\zeta({\bf x},t)=\int d^3q\left[e^{i{\bf q}\cdot{\bf x}}\alpha({\bf q})\zeta_q(t)+
e^{-i{\bf q}\cdot{\bf x}}\alpha^*({\bf q})\zeta^*_q(t)\right]\;,
\end{equation} 
\begin{equation} 
\gamma_{ij}({\bf x},t)=\int d^3q\sum_\lambda\left[e^{i{\bf q}\cdot{\bf x}}e_{ij}(\hat{q},\lambda)\alpha({\bf q},\lambda)\gamma_q(t)+
e^{-i{\bf q}\cdot{\bf x}}e^*_{ij}(\hat{q},\lambda)\alpha^*({\bf q},\lambda)\gamma^*_q(t)\right]\;,
\end{equation}
\begin{equation} 
\varphi_n({\bf x},t)=\int d^3q\left[e^{i{\bf q}\cdot{\bf x}}\alpha({\bf q},n)\varphi_q(t)+
e^{-i{\bf q}\cdot{\bf x}}\alpha^*({\bf q},n)\varphi^*_q(t)\right]\;,
\end{equation}
where $\lambda=\pm 2$ is a helicity index and $e_{ij}(\hat{q},\lambda)$ is a polarization tensor, while $\alpha({\bf q})$,  $\alpha({\bf q},\lambda)$, and $\alpha({\bf q},n)$ are conventionally normalized  annihilation operators, satisfying the usual commutation relations
\begin{equation} 
\Big[\alpha({\bf q})\,,\,\alpha^*({\bf q}')\Big]=\delta^3\Big({\bf q}-{\bf q}'\Big)\;,~~~~~~
\Big[\alpha({\bf q})\,,\,\alpha({\bf q}')\Big]=0\;.
\end{equation}
\begin{equation} 
\Big[\alpha({\bf q},\lambda)\,,\,\alpha^*({\bf q}',\lambda')\Big]=\delta_{\lambda\lambda'}\delta^3\Big({\bf q}-{\bf q}'\Big)\;,~~~~~~
\Big[\alpha({\bf q},\lambda)\,,\,\alpha({\bf q}',\lambda')\Big]=0\;,
\end{equation} 
and 
\begin{equation}
\Big[\alpha({\bf q},n)\,,\,\alpha^*({\bf q}',n')\Big]=\delta_{nn'}\delta^3\Big({\bf q}-{\bf q}'\Big)\;,~~~~~~
\Big[\alpha({\bf q},n)\,,\,\alpha({\bf q}',n')\Big]=0\;,
\end{equation}
The expectation value in Eq.~(7) is assumed to be taken in a ``Bunch--Davies'' vacuum annihilated by these annihilation operators.
Also, $\zeta_q(t)$, $\gamma_q(t)$, and $\varphi_q(t)$ are suitably normalized 
positive-frequency solutions of the wave equations
\begin{equation} 
\frac{d^2\zeta_q}{d t^2} +\left[3H+\frac{d \ln\epsilon}{dt}\right]\frac{d\zeta_q }{d t}+(q/a)^2\zeta_q=0\;,
\end{equation} 
\begin{equation} 
\frac{d^2\gamma_q}{d t^2} +3H\frac{d\gamma_q }{d t}+(q/a)^2\gamma_q=0\;,
\end{equation} 
\begin{equation}
\frac{d^2\varphi_q}{d t^2}+3H\frac{d\varphi_q}{d t}+(q/a)^2\varphi_q=0\;,
\end{equation}
where $H\equiv \dot{a}/a$ and $\epsilon\equiv -\dot{H}/H^2$.  

The functions $\zeta_q(t)$, $\gamma_q(t)$, and $\varphi_q(t)$  approach time-independent limits $\zeta^o_q$, $\gamma^o_q$, and $\varphi_q^o$ at late times during inflation, when the perturbations are far outside the horizon, with the remainders $\zeta_q(t)-\zeta_q^o$, $\gamma_q(t)-\gamma_q^o$, and $\varphi_q(t)-\varphi_q^o$
all vanishing essentially (apart from slowly varying quantities like $H$ and $\epsilon$) as 
$a^{-2}(t)$.  In consequence, $\dot{\zeta}_q(t)$, $\dot{\gamma}_q(t)$, and $\dot{\varphi}_q(t)$ all vanish at late times like $a^{-2}(t)$.  Also, as shown in reference [2], the commutator of any two interaction-picture fields at  times $t_1$, $t_2$ during inflation but long after horizon exit goes essentially as $a^{-3}(t)$, with $t$ either $t_1$ or $t_2$ or some weighted average of the times between $t_1$ and $t_2$.  The same is true for the commutator of a field and a field time derivative, while the commutator of two field time derivatives goes as $a^{-5}(t)$.\footnote{In this counting of powers of $a(t)$, we are tacitly assuming that the time dependence can be evaluated before integrating over momenta, and will not be altered when the momentum integrals are done.  This is based on the expectation that the counterterms introduced to eliminate ultraviolet divergences in flat space will suppress the contributions of large internal momenta even in an inflating spacetime.  As discussed in reference [2], this expectation is not fulfilled for arbitrary choices of the operators whose correlation functions are to be calculated.  It is necessary to consider only correlation functions of ``renormalized'' operators, for which large internal momenta {em are} suppressed.  More work needs to be done to see how to construct appropriate renormalized operators.} 

This asymptotic properties of the commutators can be seen by noting that 
\begin{eqnarray}
&&[\zeta({\bf x},t)\,,\,\zeta({\bf y},t')]=\int d^3q\;e^{i{\bf q}\cdot({\bf x}-{\bf y})}{\rm Im}\Big[\zeta_q(t)\zeta_q^*(t')\Big]\\
&&[\gamma({\bf x},t)\,,\,\gamma({\bf y},t')]=\int d^3q\;e^{i{\bf q}\cdot({\bf x}-{\bf y})}{\rm Im}\Big[\gamma_q(t)\gamma_q^*(t')\Big]\\
&&[\varphi({\bf x},t)\,,\,\varphi({\bf y},t')]=\int d^3q\;e^{i{\bf q}\cdot({\bf x}-{\bf y})}{\rm Im}\Big[\varphi_q(t)\varphi_q^*(t')\Big]\;.
\end{eqnarray}
The general solutions of Eqs.~(14)--(16) are each linear combinations with complex coefficients of two independent {\em real} solutions, one of which goes at late times as a constant plus terms of order $a^{-2}$, while the other goes as $a^{-3}$, so the imaginary parts in Eqs.(17)--(19) arise only from the interference of the two independent real solutions, which goes as $a^{-3}$.  Likewise the derivatives of these imaginary parts with respect to either $t$ or $t'$ also goes essentially as $a^{-3}$, because the derivative may act on the solution that already goes as $a^{-3}$, but the derivative with respect to both $t$ and $t'$ goes as $a^{-5}$, because both of the independent real solutions are differentiated.

We are interested in the behavior of the correlation function (7) at late times $t$, when the perturbations are far outside the horizon.  Inspection of the Lagrangian density (3) shows that no term has more than 3 factors of $a(t)$.  According to Eq.~(7), there are just as many commutators as interactions, and each commutator provides at least 3 factors of $1/a(t)$ at late times, so the total number of factors of $a(t)$ at late times in the integrals over time or in any subintegration is at most zero.  With zero factors of $a(t)$ the integrand can still grow like a power of $t$, which is more or less the same as a power of $\ln a(t)$[4], but it cannot grow like a power of $a(t)$, and therefore (without $\approx 10^{12}$ $e$-foldings) it cannot become large at late times.  

Indeed, since time derivatives of fields go like $a^{-3}(t)$, and commutators of time derivatives of fields with each other go like $a^{-5}(t)$, the integrand will go like a negative power of $a(t)$ 
if any interaction has less than 3 explicit factors of $a(t)$, or if the time derivative of a field in any interaction does not appear in a commutator, or appears in a commutator with another time derivative.  It is therefore only a very limited set of terms in the perturbation series that can contribute to a logarithmic growth of the integrand at late times.

These conclusions would not be altered by the inclusion of higher-derivative terms in the action.  Each pair of space derivatives is accompanied with a factor $g^{ij}\propto a^{-2}$, while field time derivatives of any order vanish at late times at least as fast as $ a^{-2}$.  

It remains to consider the inclusion of other kinds of fields, but first we must say a word about the effect of scalar field masses.

\begin{center}
{\bf III. MASSES}
\end{center}

In the foregoing section we have treated the scalar fields (aside from a single inflaton field whose fluctuations can be eliminated by a gauge choice) as if they were all massless, with any possible scalar mass terms in the Lagrangian implicitly included as just additional possible terms in the potential $V(\varphi)$.  As we have seen, when treated perturbatively such a term can at most introduce powers of $\ln a$ in the late-time behavior of the integrand for cosmological correlation functions.   But a mass $m$ cannot be treated as a perturbation over time intervals $t$ for which $mt\gg 1$, and in this case the powers of $\ln a$ can add up to effects that materially change the late-time behavior of the integrand, requiring a separate treatment of mass effects.

If a scalar mass $m$ is sufficiently large compared with the expansion rate $H$, then it produces oscillations in the integrand at late times, which suppresses the contribution of any times later than $1/m$.  For $m\gg H$, the correlation function is therefore dominated by times in the era of horizon exit.  But for $m < H$, a more detailed analysis is required.  

We can get a good idea of what happens in these two cases by considering the simple example of a purely exponential expansion, $a\propto e^{Ht}$, with $H$ constant.  The wave equation for any one scalar field of mass $m$ is 
\begin{equation}
\frac{d^2\varphi_q}{d t^2}+3H\frac{d\varphi_q}{d t}+\left[m^2+(q/a)^2\right]\varphi_q=0\;,
\end{equation}
For $H$ constant, the solutions for $q/aH\ll 1$ are
\begin{equation}
\varphi_q\rightarrow C_q a^{\lambda_+} \left[1+O\left(\frac{q}{aH}\right)^2\right]+
D_q a^{\lambda_-} \left[1+O\left(\frac{q}{aH}\right)^2\right]\;,
\end{equation}
where
\begin{equation}
\lambda_\pm=-\frac{3}{2}\pm \sqrt{\frac{9}{4}-\frac{m^2}{H^2}}\;,
\end{equation}
and $C_q$ and $D_q$ are complex constants determined by matching this solution to solutions before horizon exit.
For $m>3H/2$ the exponents $\lambda_\pm$ are complex conjugates, so as mentioned above, the oscillations of the wave functions suppress the contribution of late times.  

For $m<3H/2$, the $\lambda_\pm$ are real, with 
$$-3/2<\lambda_+<0\;,~~~~-3<\lambda_-<-3/2\;.$$  Each scalar field factor in the Lagrangian thus contributes a factor of $ a^{\lambda_+}(t)$ at late times, and as long as $q/aH\ll \sqrt{\lambda_+}$, the time derivative of a scalar field will contribute the same factor.  On the other hand, commutators of scalar fields and/or scalar time derivatives  contribute factors $a(t)^{\lambda_++\lambda_-}=a^{-3}(t)$, since the commutators can arise only from an interference between the two terms in Eq.~(21).  Once again, with no more than 3 powers of $a(t)$ in each interaction, and with just as many commutators as there are interactions, the total number of factors of $a(t)$ in the integrands for correlation functions cannot be greater than zero.  Furthermore, except for trivial diagrams in which every vertex has just two lines attached, since each commutator involves just two fields, there must be fields that are {\em not} in  commutators.  These contribute additional factors of $a(t)^{\lambda_+}$ to the integrand, and since $\lambda_+<0$, the integrand will be exponentially damped at late times, and the correlation functions will depend only on the behavior of the fields near horizon exit.

\begin{center}
{\bf IV. VECTOR FIELDS}
\end{center}

Next consider a massless vector field, given (in temporal gauge) in the interaction picture by
\begin{equation}
A_i({\bf x},t)=\sum_\lambda\int d^3q \left[e^{i{\bf q}\cdot {\bf x}}\epsilon_i(\hat{q},\lambda)\alpha({\bf q},\lambda)u_q(t)+
e^{-i{\bf q}\cdot {\bf x}}\epsilon_i(\hat{q},\lambda)\alpha^*({\bf q},\lambda)u^*_q(t)\right]\;,
\end{equation}
where here $\epsilon_i(\hat{q},\lambda)$ and $\alpha({\bf q},\lambda)$ are the polarization vectors and annihilation operators for massless  particles of helicity $\lambda=\pm 1$, and 
$ u_q(t)$ is a suitably normalized solution of the wave equation
\begin{equation}
\frac{d}{dt}\left(a(t)\frac{d}{dt}u_q(t)\right)+\frac{q^2}{a(t)}
u_q(t)=0
\end{equation}
The commutator of two vector fields at unequal times is then
\begin{eqnarray}
[A_i({\bf x},t)\,,\, A_j({\bf x}',t')]&=&\int d^3q \Big(\delta_{ij}-\hat{q}_i\hat{q}_j\Big)
\,\exp\Big(i{\bf q}\cdot({\bf x}-{\bf x}')\Big)\nonumber\\&&\times \Big(u_q(t)\,u^*_q(t')-u_q(t')\,u_q^*(t)\Big)\;.
\end{eqnarray}
Now, the general solution of the wave equation (24) here takes the simple form
\begin{equation}
u_q(t)= C_q\,\cos q\tau +D_q\,\sin q\tau\;,
\end{equation}
with $C_q$ and $D_q$ complex constants, and as usual
\begin{equation}\tau\equiv \int^\infty_t \frac{dt'}{a(t')}
\;.
\end{equation}
We see that at late times, where $\tau\rightarrow 0$, $u_q(t)$ approaches a constant, while $\dot{u}_q(t)$ goes essentially as $1/a(t)$.  Also,  
\begin{equation}
u_q(t)\,u^*_q(t')-u_q(t')\,u_q^*(t)=2i{\rm Im} \Big(C_qD_q^*\Big)\,\sin \Big(q(\tau-\tau')\Big)
\end{equation}
so at late times $u_q(t)u_q^*(t')- u_q(t')u_q^*(t)$ and
$u_q(t)\dot{u}_q^*(t')- \dot{u}_q(t')u_q^*(t)$  go essentially as $1/a$, while $\dot{u}_q(t)\dot{u}_q^*(t')- \dot{u}_q(t')\dot{u}_q^*(t)$ goes to zero even faster, as $1/a^3$.    There are just as many commutators as there are interaction vertices, so if a term in the integrand involves a set of interactions $H_s$ with $A_s$ explicit factors of $a$, the integrand will contain altogether a number of factors of $a$ bounded by
\begin{equation}
\#\leq \sum_s[A_s-1]\;.
\end{equation}
Because of the vector nature of the field, the maximum number of explicit factors of $a(t)$ in any interaction is $3-2=1$.  For instance, in temporal 
gauge the electromagnetic interaction of a charged scalar field $\varphi$ is
\begin{equation}
a^3a^{-2}\Big[ieA_i\Big(\varphi^*\partial_i\varphi-\varphi\partial_i\varphi^*\Big)-e^2A_iA_i\varphi^*\varphi\Big]\;,
\end{equation}
with the factor $a^3$ coming from the metric determinant and the factor $a^{-2}$ coming from $g^{ij}$.
So again the maximum number of factors of $a$ in the integrand is zero, giving an integrand that grows at most like a power of $\ln a$.  Derivative interactions of the vector field behave even better, because time-derivatives of vector fields give extra factors of $1/a$, while pairs of space-derivatives are accompanied with factors $g_{ij}\propto a^{-2}$.   For non-Abelian gauge fields $A_{\alpha\mu}$, there are self-interactions 
\begin{equation}
-a^3a^{-4}\left[C_{\alpha\beta\gamma}\partial_iA_{\alpha j}A_{\beta i}A_{\gamma j}+\frac{1}{4}C_{\alpha\beta\gamma}C_{\alpha\delta\epsilon}A_{i\beta}A_{j\gamma}A_{i\delta}A_{j\epsilon}\right]\;,
\end{equation}
where $C_{\alpha\beta\gamma}$ is a structure constant.  The four factors of $1/a$ appear here because the interaction involves two contractions of space indices.  Each such interaction contributes a factor $a^{-2}$ to the integrand, suppressing the contribution of late times.

\begin{center}
{\bf V. DIRAC FIELDS}
\end{center}

A Dirac field of mass $m$ in the interaction picture involves a wave function $\psi_q(t)$ that satisfies the wave equation
\begin{equation}
\frac{d}{dt}\psi_q+\frac{3H}{2}\psi_q+i a^{-1}\gamma_0 \gamma_iq_i\psi_q+\gamma_0m\psi_q=0\;.
\end{equation}
Hence for wave numbers far outside the horizon, the Dirac wave function has the asymptotic limit
\begin{equation}
\psi_q(t)\propto e^{-\gamma_0mt}a^{-3/2}(t)\;.
\end{equation}
The matrix $\gamma_0$ has eigenvalues $\pm i$, so the factor $ e^{-\gamma_0mt}$ produces an oscillation, which does not affect bilinears like $\bar{\psi}\psi$ or $\bar{\psi}\gamma_0\psi$, but does produce an oscillation in bilinears like $\bar{\psi}\gamma_i\psi$, which suppresses the late-time contribution of interactions containing such bilinears.  Even apart from this factor (as for instance for $m=0$), every bilinear combination of $\psi$ and $\bar{\psi}$ is suppressed by a factor $a^{-3}$ produced by the factor $a^{-3/2}$ in Eq.~(33).  This in itself cancels the $a^3$ factor from the metric determinant, so that no positive powers of $a(t)$ can be produced by any interaction involving Dirac fields.

\begin{center}
{\bf VI. AFTERTHOUGHT}
\end{center}

In generic theories  the $N$ integrals over time in $N$-th order perturbation theory will yield correlation functions at time $t$ that grow as $(\ln a(t))^N$.  Such a power series in $\ln a(t)$ can easily add up to a time dependence that grows like a power of $a(t)$, or even more dramatically.  As everyone knows, the series of powers of the logarithm of energy encountered in various flat-space theories such as quantum chromodynamics can be summed by the method of the renormalization group.  It will be interesting to see if the power series in $\ln a(t)$ encountered in calculating cosmological correlation functions at time $t$,  though arising here in a very different way, can be summed by similar methods.

\begin{center}
{\bf ACKNOWLEDGMENTS}
\end{center}

For helpful conversations I am grateful to K. Chaicherdsakul.  This material is based upon work supported by the National Science Foundation under Grants Nos. PHY-0071512 and PHY-0455649 and with support from The Robert A. Welch Foundation, Grant No. F-0014, and also grant support from the US Navy, Office of Naval Research, Grant Nos. N00014-03-1-0639 and N00014-04-1-0336, Quantum Optics Initiative.

\begin{center}
{\bf REFERENCES}
\end{center}
\nopagebreak

\begin{enumerate}

\item J. Maldacena, J. High Energy Phys. {\bf 0305}, 013 (2003) (astro-ph/0210603).  For other work on this problem, see A. Gangui, F. Lucchin, S. Matarrese,and S. Mollerach, Astrophys. J. {\bf 430}, 447 (1994) (astro-ph/9312033); P. Creminelli, J. Cosm. Astropart. Phys. {\bf 0310}, 003 (3002) (astro-ph/0306122); P. Creminelli and M. Zaldarriaga, J. Cosm. Astropart. Phys. {\bf 0410}, 006 (2004) (astro-ph/0407059); G. I. Rigopoulos, E.P.S. Shellard, and B.J.W. van Tent, Phys. Rev. D {\bf 72}, 08357 (2005) (astro-ph/0410486); F. Bernardeau, T. Brunier, and J-P. Uzan, Phys. Rev. D {\bf 69}, 063520 (2004). For a review, see N. Bartolo, E. Komatsu, S. Matarrese, and A. Riotto, Phys. Rept. {\bf 402}, 103 (2004) (astro-ph/0406398).

\item S. Weinberg, Phys. Rev. D {\bf 72}, 043514 (2005) 
(hep-th/0506236).

\item R. S. Arnowitt, S. Deser, and C. W. Misner, in {\em Gravitation: An Introduction to Current Research}, ed. L. Witten (Wiley, New York, 1962): 227; now available as gr-qc/0405109.

\item A $\log a(t)$ dependence has been found in different contexts by Woodard and his collaborators; see e. g.  N. C. Tsamis and R. Woodard, Ann. Phys. {\bf 238}, 1 (1995); {\bf 253}, 1 (1997); N. C. Tsamis and R. Woodard, Phys. Lett. {\bf B426}, 21 (1998); V. K. Onemli and R. P. Woodard, Class. Quant. Grav. {\bf 19}, 4607 (2002); T. Prokopec, O. Tornkvist, and R. P. Woodard, Ann. Phys. {\bf 303}, 251 (2003); T. Prokopec and R. P. Woodard, JHEP {\bf 0310}, 059 (2003); V. K. Onemli and R. P. Woodard, Phys. Rev. D {\bf 70}, 107301 (2004); T. Brunier, V.K. Onemli, and R. P. Woodard, Class. Quant. Grav. {\bf 22}, 59 (2005).

\end{enumerate}

\end{document}